\documentclass[letterpaper,twocolumn,10pt]{article}

\usepackage[english]{babel} 
\usepackage{amsmath} 
\usepackage{graphics}
\usepackage{graphicx} 
\usepackage{cite} 
\usepackage[hyphens]{url}
\usepackage{authblk}
\usepackage{todonotes}
\usepackage{caption}
\usepackage{subcaption}
\usepackage{flushend}
\usepackage{listings}

\newcommand{\approach}{{\sc IdentityMailer}}
\title{That Ain't You: Blocking Spearphishing Emails \\ Before They Are Sent} 

\author[$\S$]{Gianluca Stringhini}
\author[${\ddag}$]{Olivier Thonnard}
\affil[$\S$]{University College London \hspace{7em}
\textsuperscript{${\ddag}$} Amadeus\authorcr
\hspace{1in}\texttt{g.stringhini@ucl.ac.uk}
\hspace{0.5in}\texttt{olivier.thonnard@amadeus.com}
\authorcr 
}
\date{}
\begin{document}

\maketitle 
\thispagestyle{empty}

\begin{abstract}

One of the ways in which attackers try to steal sensitive information from corporations is by
sending \emph{spearphishing} emails. This type of emails typically appear to be sent by one of the victim's
coworkers, but have instead been crafted by an attacker. 
A particularly insidious type of spearphishing emails are the ones that do not only
claim to come from a trusted party, but were actually sent from that party's legitimate email account
that was compromised in the first place.  
In this paper, we propose a radical change of focus in the
techniques used for detecting such malicious emails: instead of looking for
particular features that are indicative of attack emails, we look for possible indicators of 
\emph{impersonation} of the legitimate owners. 
We present \approach{}, a system that validates the authorship of emails
by learning the typical email-sending behavior of users over time, 
and comparing any subsequent email sent from their accounts against this model. 
Our experiments on real world e-mail datasets demonstrate that our system can effectively block
advanced email attacks sent from \emph{genuine} email accounts, which traditional protection systems are unable to detect. 
Moreover, we show that it is resilient to an attacker willing to evade the system. 
To the best of our knowledge, \approach{} is the first system able to identify spearphishing emails 
that are sent \emph{from within an organization}, by a skilled attacker having access 
to a compromised email account.

\end{abstract}
\section{Introduction}

Companies and organizations are constantly under attack by cybercriminals trying to
infiltrate corporate networks with the ultimate goal of stealing sensitive
information from the company.
Such an attack is often started by sending a \emph{spearphishing} email.
Attackers can breach into a company's network in many ways, for example by leveraging 
advanced malware schemes~\cite{nightdragon}. After entering the network,
attackers will perform additional activities aimed at gaining access to more
computers in the network, until they are able to reach the sensitive information
that they are looking for. This process is called \emph{lateral movement}.
Attackers typically infiltrate a corporate network, gain access to
internal machines within a company and acquire sensitive information by sending
\emph{spearphishing} emails. 
In a spearphishing attack an email is crafted 
and sent to a specific person within a company, with
the goal of infecting her machine with an unknown piece of malware, luring her to hand out access
credentials, or to provide sensitive information. Recent research showed that
spearphishing is a real threat, and that companies are constantly targeted
by this type of attack~\cite{thonnard2012industrial}. 

\noindent\textbf{Spearphishing is not spam.} 
While they may share a few common characteristics, it is important to note that spearphishing is still very different from traditional email spam. 
In most cases, spearphishing emails appear to be coming from accounts within the
same company or from a trusted party, to avoid raising suspicion by the victim~\cite{trend2012spearphishing}.
This can be done in a trivial way, by forging the \texttt{From:} field in the attack email.
However, in more sophisticated attacks, the malicious emails are actually sent from a \emph{legitimate}
employee's email account whose machine has been compromised, or whose 
credentials have been previously stolen by the attacker~\cite{tibet}. 
From the attacker's perspective, this modus operandi presents two key advantages. First, it leverages
a user's social connections: previous research showed that users are more
likely to fall for scams if the malicious message is sent by somebody they trust~\cite{jagatic2007social}. 
Secondly, it circumvents existing detection systems, which are typically based on anti-spam techniques. This happens for two reasons:
first, the content of spearphishing emails looks in many cases 
completely legitimate and it does not contain any words that are
indicative of spam, since the goal is to make it resemble typical business emails. Second, if an
email impersonating one of the company's employees comes
from that person's computer, which has been compromised, then origin-based detection
techniques, such as IP reputation, become useless.
Secondly, it circumvents all IP and origin-based blacklisting systems, as well as email sender or domain verification systems such as \emph{Sender Policy Framework} (SPF) and \emph{DomainKeys Identified Mail} (DKIM)~\cite{Leiba2007,spf:06}, since the email is sent from a genuine email account.

\noindent\textbf{A new paradigm for fighting targeted attack emails.}
Given how different spearphishing emails are compared to 
traditional spam and phishing emails, we propose a \emph{paradigm shift} in detection
approaches to fight this threat, and present \approach{}, a system to detect and
block spearphishing emails sent from compromised accounts. 
Instead of looking for signs of maliciousness in emails (such as words that are indicative of illicit content, phishy-looking content, or suspicious
origin), \approach{} determines whether an email was actually written by the author
that it claims to come from. In other words, we try to automatically validate the genuineness of the email authorship.
Our approach is based on a simple, yet
effective observation: most users develop habits when sending emails. These habits
include frequent interactions with specific people, sending emails at specific
hours of the day, and using certain greetings, closing statements, and modal words in their emails. 
The core of \approach{} consists in building a user profile reflecting her email-sending behavior.
When a user's account gets compromised, the attack
emails that are sent from this account are likely to show differences from the behavioral profile of the genuine user. 

Behavioral anomalies can be very evident or more subtle. An
example of a ``noisy'' attack is a worm that sends an email to the entire address
book of a user~\cite{worm}, which is a behavior that typical users do not show.
In more realistic scenarios, attackers might try to mimic the typical
behavior of the person they are impersonating in their emails. 
What they could do is sending emails
only at hours in which the user is typically sending them, and only to people she
frequently interacts with, or even imitate the user's writing style. 

To make it more difficult for attackers to successfully evade our system, 
\approach{} builds the email-sending behavioral profile for a particular user by
leveraging both the emails previously sent by that specific user and a set of emails
that other users in the organization authored. In a nutshell, \approach{}
compares the emails written by the user to the ones written by everybody else
and extracts those characteristics that are the most representative of the
user's behavior. If common characteristics are shared by multiple users,
however,
these will be de-emphasized because they are not specific to a particular user.
For example, certain functional words only used by a given user 
(and rarely by others) would model her behavior well.

When an attacker tries to learn a victim's sending behavior to mimic it in his
attack emails, he only has access to that user's emails (since he compromised
her account or personal machine). It is unlikely, however, that he has access to the ones authored by all other users within
the company -- besides the few emails exchanged by the victim and the coworkers he/she is interacting with. Therefore, all the attacker can do is learning the most common habits of
the user (such as the email address that is more frequently contacted, and at
what time the user generally sends emails), but he has no guarantee that those
traits are actually representative of the victim's behavior. 

\noindent\textbf{Working on the sending end.}
Traditional anti-spam systems work on the receiving end of the
email process. This means that they analyze incoming emails, and establish
whether they are legitimate or malicious. This approach 
is very effective in general, but it has many drawbacks in our specific
case. First of all, the analysis that can be performed on incoming emails
has to be lightweight, due to the large amount of emails, mostly malicious, that
mail servers receive~\cite{Taylor2006}. As a second drawback, learning the
typical behavior of a user on the receiving end has the problem that a mail
server only has visibility of the emails that are exchanged between that user
and people whose mailboxes are hosted on the server. Therefore, a behavioral
profile built from these emails might not be representative enough to correctly
model the sending habits of users.

Because of the aforementioned problems, 
we propose to perform the analysis when emails are sent, before they are forwarded to the outgoing SMTP server. 
Our approach builds a behavioral profile based on the emails that a user sent in the 
past (and a set of emails authored by the other people in the organization). 
Then, every time an email is sent by
that account, our approach checks if this email matches the profile learned for the
real account's owner. If the email does not match the learned profile, we consider
it anomalous. The account might have been compromised, and the email might
actually be an attack attempt. However, the anomaly might also be a false
positive. Perhaps the user is working on a deadline, and is sending emails late at
night, or is sending a personal email, and using a colloquial language, while
the account is primarily used to send work-related emails. 

False positives are a
big problem in traditional anti-spam systems, because they annoy users in the best case,
and they prevent them from receiving important emails in the worst case.
Luckily, the fact that our approach operates on the sending side of the email
process comes to our aid. Any time an email is flagged as anomalous, we can start a 
process to verify the identity of a user. In our design, this process would
include sending a confirmation code to a device owned by the user, as part of a two-factor authentication scheme~\cite{aloul2009two}. 
If the user correctly inputs the confirmation code when asked we consider
the anomaly as a false positive, and the email is forwarded. In addition, we update
the user's behavioral profile to include this particular email, so that we will we avoid similar
false positives in the future. If the user fails at 
solving the challenge, however, we consider the email as a possible attack, and we
discard it. We acknowledge that having to go through an identity-verification process can be annoying for users. However, we
think that having users confirm their identity once in a while is a fair price to pay 
to protect a company against advanced email attacks, as long as the verifications are rare enough (for example, one in every 30 emails on average).

We tested \approach{} on a large set
of publicly-available emails, and on real world data sets made of malicious targeted emails sent
to the customers of a large security company.

In summary, this paper makes the following contributions: 
\begin{itemize} 
\item We propose a new approach to detect spearphishing emails sent from compromised accounts: instead of 
looking for signs of maliciousness, we introduce a set of features that are
representative of the email-sending behavior of a user, and propose a method to
check emails against the learned sending behavior (\emph{i.e.}, email authorship validation).  
\item We implemented the approach in a system called \approach{}, which was tested against a large
dataset of publicly-available emails, as well as other real-world datasets of targeted attack emails. 
Our experimental results show that our approach can effectively block spearphishing emails that 
state-of-the-art systems failed to detect with a detection rate above 
90\% for users with a \emph{sent} history of only 1,000 emails. In contrast, existing systems that look for
 signs of maliciousness are failing to detect most of these advanced email attacks.
\item 
  We show that having access to the emails sent by the victim is not sufficient to evade \approach{}, and that
imitating the most common characteristics present in those emails can even augment the chances of being detected by the system. 
\end{itemize}

\section{Threat Model}

Spearphishing can be broadly defined as targeted phishing. In traditional
phishing emails, attackers pretend to be operators from online services such as
Online Social Networks or Online Banking Portals, and lure their victim into
inserting their credentials on a fake web page that reproduces the home page of
that online service. 
An example of phishing email is shown in
Figure~\ref{fig:phishing}. 
A spearphishing email is a type of phishing email for which
the attacker collected some information about the victim, and included it in the
attack email to make it more believable~\cite{parmar2012protecting}.

A spearphishing attack can be more or less sophisticated. An example of an
attack with a low sophistication is a password reset request for a corporate
service sent to all the employees of a company. In this case, the only
difference with a traditional phishing email is that,
unlike phishing attempts against common online services, all
people receiving the malicious email supposedly have an account on the corporate
service. A more sophisticated spearphishing attack is one that targets a
specific person within a company, and leverages personal information about that
person in order to make the email look even more convincing.
Attacks of such sophistication can usually be detected by variations of
traditional anti-spam and anti-phishing techniques. For instance, the attack email might be coming from an IP address that does not belong to
the online service that it is allegedly coming from; similarly, the URL of the link
in the phishing email might not hosted on the domain of the online service. Systems
that look at the source IP address of an
email~\cite{Ramachandran:07:behavioral,Sinha:2010:ISB,spamhaus} or at the actual phishing web
page~\cite{zhang2007cantina,fette2007learning} are able, in general, to detect such attacks.

\begin{figure}[t]
\begin{lstlisting}[frame=trbl]
Received: from [FOREIGN IP]
From: <support@site.com>
To: <victim@company.com>

Dear user,
Your account has been hacked. 
Please reset your password 
<a href="http://fakesite.org">
here </a>
\end{lstlisting}
\caption{Example of a traditional phishing email. The source IP address does not
belong to the online service that the email claims to come from, and the URL points to
a phony web site.}
\label{fig:phishing}
\end{figure}

\begin{figure}[t]
\begin{lstlisting}[frame=trbl]
Received: from [COMPANY IP]
From: <manager@company.com>
To: <victim@company.com>

Dear <victim>,
As discussed during our meeting, 
please send me the latest report. 
The template to use is attached.

Thanks,
<manager>

\end{lstlisting}
\caption{Example of advanced spearphishing email. The sender account has been
compromised, therefore the sender IP address is not anomalous. Also, the content of
the email does not look anomalous: the language used looks like regular business emails.}
\label{fig:spearphishing}
\end{figure}

Although existing anti-spam and anti-phishing techniques can be adapted to detect
certain types of spearphishing attacks, as the attacker gets even more sophisticated, these
techniques become inadequate in fighting this threat. Consider, for example, the
spearphishing email in Figure~\ref{fig:spearphishing}. In this case, the
attacker managed to compromise the email account of one of the company's managers, and
is using it to send an email to a member of his team. In the attack email,
the attacker asks the victim to send his boss a copy of the latest report. The
victim is very likely to fall for this attack because there is nothing
indicative of a malicious email: the email is coming from the email account of
the manager, and therefore the IP address is the correct one; similarly, the
language of the email is not suspicious at all, because the language used is typical of
regular business emails. In addition, since the attacker has access to the
manager's email account, he can retrieve the report directly from there, once
the victim sends it, and does not need to include the link to an external web
server in the spearphishing email. From the perspective of traditional anti-spam
and anti-phishing techniques, the email in Figure~\ref{fig:spearphishing} looks perfectly 
authentic, and has no indicators for being flagged as anomalous. 

In this paper, we propose a technique that can detect sophisticated
spearphishing emails that do not present any signs of maliciousness. 
As we said, our approach works as follows: first, we learn the typical behavior of a user
based on his sent email history; then, we compare each new email sent from that user account to 
determine if it does match the learned behavior. In the following sections, we describe our approach in
more detail.

\section{Behavioral Profiles}

In the very first stage, \approach{} must accurately learn and model the email-sending behavior of a user,
as it will enable us to perform better detection of anomalous emails in a later stage. However, 
defining user-specific traits that best distinguish a user's
sending behavior is not trivial.
To determine these traits, \approach{} requires two datasets: a set
$\textbf{M}_u$ of emails written by a user $U$ and a set $\textbf{M}_o$ of
legitimate emails
written by other people. 
By comparing the emails in
$\textbf{M}_u$ to the ones in $\textbf{M}_o$, we can extract the distinguishing
characteristics of the email-sending behavior of $U$.

$\textbf{M}_o$ should be composed of both emails sent by people working in
the same organization as $U$, as well as of emails written by people who are
completely unrelated to $U$. 
As we will explain later, the privacy concerns of our approach are minimal,
because we do not save the full email, but only a feature vector associated to
it.
On one side, having $\textbf{M}_o$ built from
the emails sent by the users working in the same organization as $U$ helps in
giving less importance
to common characteristics shared by coworkers.
For example, if no user in the organization ever sends emails on Sundays, it is
less peculiar if the user follows this trend. On the other hand, having emails sent
by users who are completely unrelated to $U$ in $\textbf{M}_o$ helps giving to the
model examples of behavioral characteristics that are uncommon in the organization, but
common outside of it. We provide a more detailed description on how we build $\textbf{M}_o$ in Section~\ref{sec:building}.
By using only legitimate emails to build our
behavioral profiles, \approach{} does not need to have ever observed any attack email to
perform detection, similarly to what happens with traditional anomaly-detection
systems.
This is important, because it makes our approach independent from specific attack schemes.

To build the email-sending behavioral profile for a user, we proceed in two steps. First, we
extract a number of features for each email  in $\textbf{M}_u$ and $\textbf{M}_o$. 
As a second step, we leverage these feature vectors to build a classification model,
which represents the actual behavioral profile. We use this profile to analyze
any email written by the user, and determine whether it was really written by
that user.

\subsection{Extracting Behavioral\\ Email Features} \label{sec:features}

We define three
classes of email features, which pertain to: \emph{writing habits}, \emph{composition habits}, and
\emph{interaction habits}. Previous research showed that authorship
identification is possible by looking at stylometry features 
(which are a
subset of what we call writing habits)
~\cite{corney2003}. However, these
approaches rely on texts of a certain length (250 words or more)~\cite{forsyth1996feature}. 
Unfortunately, as we show in
Section~\ref{sec:evaluation}, many emails are short. If our approach relied only
on the writing habits of a user (i.e., stylometry features), it would fail at detecting short attack emails. Therefore, we need
additional information to deal with such emails, as described here after. 
In the following, we describe the features that our approach uses to characterize an email.

\noindent\textbf{Writing habits.} People normally develop their own writing style.
For example, some people use certain functional words (such as
``although'') more often than others, or write dates in a certain way.
Analyzing a user's style has been used in the past to determine authorship of
texts and
emails~\cite{corney2003,abbasi2008stylometric,narayanan2012feasibility}.
Similarly, we consider a user's writing style as a strong indicator of 
email authorship. An attacker could, in principle, learn
the characteristics of his victim's style, and replicate them in the attack
emails that he sends. However, previous research showed that imitation of another
person's writing style is detectable most of the time~\cite{afroz2012detecting}.
In addition, as we will show in Section~\ref{sec:adapting}, it is difficult for an
attacker to figure out which features are the most representative of a user's
writing style.
In the following, we define a number of features that help defining a
user's writing style. 
The complete list of writing-habit
features used by \approach{} can be found in Table~\ref{sec:featureswriting}.

\begin{table*}
\begin{center}
\scalebox{0.7}{
\begin{tabular}{l l l l l l l l}
\multicolumn{3}{l}{\textbf{Character occurrence}}&\\
\multicolumn{3}{l}{Punctuation} & \multicolumn{5}{l}{. : ; , ' " ? !} \\
\multicolumn{3}{l}{Special characters} & \multicolumn{5}{l}{\% \_ \& \$ @ * $\backslash$ > < \# / -} \\
\multicolumn{3}{l}{Parenthesis} & \multicolumn{5}{l}{$()$ $[]$ $\{\}$} \\
\multicolumn{3}{l}{Ordinals} & \multicolumn{5}{l}{0123456789} \\
\multicolumn{3}{l}{Capital letters} & \multicolumn{5}{l}{ABCDEFGHIJKLMNOPQURSTVWXYZ} \\
\multicolumn{3}{l}{\textbf{Functional words}}&\\
a & aboard & about & above & absent & according to & across & after\\
against & ahead to & all & along & alongside & amid & amidst & although\\
am & among & amongst & and & another & any & anybody & anyone \\ 
anything & are & aren't & arent & are not & around & aint & ain't \\
around & as to & as far as & as well as & as & aside from & 're &  aslant \\
astride & ah & at & athwart & atop & because & because of & been \\
before &  behind & believe & below & beneath & besides & beside & best\\
between & be & beyond & both & but & by means of & by & can not \\
cannot & can't & can & cant & can you & close to & concerning & considering \\
cause & cos & could & despite & did not & didn't & didnt & did \\
do & does & do not & don't & dont & down & due to & during \\
each other & each & either & enough & even if & even though & even & everybody \\
everyone & everything & every & except for & except & far from & few & following \\
for & from & had not & hadn't & hadnt & had & have & ahead to\\
he & herself & hers & her & himself & him & his & I believe \\
hi & hello & happy & take care & regards & please & greetings & cheers \\
I'd & if & I'll & in accordance with & in addition to & in case & in case of & including\\
in front of & in lieu of & in order that & in order to & in place of & inside of & inside & in spite of\\
instead of & instead & in to & into & in & is & it & I\\
latter & less & like & little & look & lots & many & may \\
me & mid & might not & mightn't & mightnt & might & mine & minus\\
more & most & much & must not & mustn't & must not & mustnt & must \\
myself & my & near to & near & need & neither & next to & next \\
nobody & none & no one & nope & nor & nothing & notwithstanding & no \\
now that & off & of & okay & OK & on account of & on behalf of & once\\
one another & one & on to & on top of & onto & on & opposite & or \\
others & other & ourselves & our & out from & out of & outside of & outside \\
out & over & owing to & own & past & per & please & plenty \\
plus & prior to & pursuant to & quite & regarding & regardless of & round & same \\
seem & several & shall not & shall & shan't & shant & she & should not \\
shouldn't & shouldnt & since & somebody & someone & something & somewhere & some \\
soon & so & subsequent to & such & take care & thanks & thank you & than \\
that & the & theirs & their & themselves & them & then & these \\
they & think & this & those & though & throughout & through & till \\
towards & toward & to & under & unless & unlike & until & upon \\
up & used & us & versus & via & was not & wasn't & wasnt \\
was & we & were not & weren't & werent & were & what & whatever \\
when & whereas & where & whether or not & whether & which & whichever & while \\
who & whoever & whomever & whom & whose & will & will not & with \\
within & without & won't & wont & worth & would & yes & yet \\
you & yourself & yourselves & yours & your \\
\multicolumn{3}{l}{\textbf{Special word occurrence}}&\\
Full name & date & day of the week & shortened day \\ month & shortened month & year & phone number \\ 
dollar amount & time of the day & fraction \\ 
\multicolumn{3}{l}{\textbf{Generic style characteristics}}&\\
\multicolumn{3}{l}{Emoticons} & \multicolumn{5}{l}{:) :-) :P :-P :( :-( :/ :-/} \\
\multicolumn{3}{l}{Bullets type 1} & \multicolumn{5}{l}{1) 2) 3) ...}\\
\multicolumn{3}{l}{Bullets type 2} & \multicolumn{5}{l}{1- 2- 3- ...}\\
\multicolumn{3}{l}{Bullets type 3} & \multicolumn{5}{l}{1. 2. 3. ...}\\
\multicolumn{3}{l}{Bullets type 4} & \multicolumn{5}{l}{(i) (ii) (iii) ...}\\
\multicolumn{3}{l}{Bullets type 5} & \multicolumn{5}{l}{first second third ...}\\
\multicolumn{3}{l}{Bullets type 6} & \multicolumn{5}{l}{-- ...}\\
\multicolumn{3}{l}{Programming language keywords} & \multicolumn{5}{l}{if then else while do switch case return}\\
comma separated list & oxford comma & signature &  \multicolumn{2}{l}{no space after punctuation} \\ no comma in large digits & comma in large digits  \\
\multicolumn{3}{l}{\textbf{Style metrics}}&\\
\multicolumn{2}{l}{number of paragraphs} & \multicolumn{2}{l}{sentences per paragraph} & unique words & number of words & message length \\
\multicolumn{2}{l}{number of long lines} & \multicolumn{2}{l}{number of short lines} & \multicolumn{4}{l}{frequency of words of length N (from 1 to 20)} \\
\multicolumn{2}{l}{hapax legomena} & \multicolumn{2}{l}{hapax dislegomena} & \multicolumn{2}{l}{Sichel's S measure} & \multicolumn{2}{l}{Honore's R measure} \\
\multicolumn{2}{l}{Yule's metric} & \multicolumn{2}{l}{Simpson's metric} \\
\multicolumn{3}{l}{\textbf{Context-specific words}}&\\
account & accounting & attach & attached & attachment & authorize & authorizing & bank \\ candidate & cash & capacity & consideration & contract & detail & derivative & document \\ email & e-mail & Enron & equity & fax & financial & flow & gas \\ increase & information & item & manage & market & oil & option & potential \\ recommend & reserve & review & risk & scan & section & send & sent \\ stock & supply & target & trader & trading & trip \\
\end{tabular}
}
\caption{List of writing-habit features used by \approach{}.}
\label{sec:featureswriting}
\end{center}
\end{table*}

\noindent\emph{1) Character occurrence (62 features).} These features represent how often a
character, or a set of characters, appear in the email text. Given a set of
characters $\textbf{C}$ and an email text $M$, we define the character
occurrence of $\textbf{C}$ in $M$ $o_c$ as the number of times that any of the
characters in $\textbf{C}$ occur in $M$, divided by the length of
$M$. Examples of character occurrence features include the frequency of
alphabetical letters (such as ``a''), the frequency of certain punctuation signs
(such as ``;''), and the frequency of sets of characters (such as capital
letters or cardinal numbers).

\noindent\emph{2) Functional word occurrence (344 features).} These features represent how often
the person uses specific functional words. We define as functional words those
words that do not serve to express content, but instead are used to express
grammatical relationships with other words within a sentence. These include
adverbs (such as ``when''), auxiliary verbs (such as ``is''), and prepositions
(such as ``for''). Some of these features are useful to determine whether a user
uses certain functional words in their extended or shortened form, and to what
extent (for example, whether she usually uses ``don't'' instead of ``do not'').
Given a word $FW$ and a set of words $\textbf{W}_m$ in an email, we calculate the
word occurrence $o_{fw}$ in $\textbf{W}_m$ as the number of times $FW$ occurs in the
email, divided by the size of $\textbf{W}_m$.

\noindent\emph{3) Special word occurrence (11 features).} These features represent how often a
user uses certain ``special'' words in her emails. Special words include full
names, dates, and acronyms. Given a regular expression $R_{sw}$ representing the
special word, an email $M$, and a set $\textbf{W}_m$ containing the
words in $M$, we calculate the special word occurrence $o_{sw}$ of
$R_{sw}$ as the number of matches in $M$ for $R_{sw}$, divided by the
size of $\textbf{W}_m$.

\noindent\emph{4) Generic style characteristics (38 features).} These features represent
generic characteristics of the style of a user. Examples include the type of
bullets that the user uses in lists (``1-'', ``1.'', or others), whether she uses a
comma as a separator for large digits or not, and whether she uses a space after
punctuation. Given a set of regular expressions $\textbf{R}_{sc}$ representing a
style characteristic, an email $M$, and a set $\textbf{W}_m$ containing
the words in $M$, we define the style characteristic $s_c$ as the
number of matches of the regular expressions in $\textbf{R}_{sc}$ in the email
$M$, divided by the size of $\textbf{W}_m$.

\noindent\emph{5) Style metrics (33 features).} These features capture information about the
style of entire emails. Some features are rather simple, such as the
number of paragraphs in the email. Others are more advanced, and depict the
expressiveness of the language used in the email. Examples are the \emph{Sichel
measure} or the \emph{Yule metric}, which describe how complex the vocabulary used by 
an author is. These metrics have been already used in previous work on
authorship identification~\cite{yule1944statistical,tweedie1998variable}.

\noindent\emph{6) Context-specific words (variable number of features).} These features look for occurrences of
words that are common in a certain industry. People working in that industry
will use them more or less frequently, depending on their role in the company
and their specific job. Examples of context-specific words for a financial institution 
include the words ``stock'', ``asset'', and ``contract.''
Given a word $W$ and a set of words $\textbf{W}_m$ in an email, we calculate the
context-specific word occurrence $o_{wcs}$ as the number of occurrences of $W$ in
$\textbf{W}_m$, divided by the size of $\textbf{W}_m$.
Context-specific words vary with the type of business that the company is doing,
and have been used in other authorship-recognition research~\cite{zheng2005framework}.
We discuss the choice of context-specific words that we used in our experiments
in Section~\ref{sec:classifieranalysis}.

\begin{table*}[t]
\begin{center}
 \begin{tabular}{l l l}
  \textbf{Message characteristics:} & has signature & has URL \\ 
  & indented lines & quoted lines \\
  & original attached & has attachment \\
  & is reply & is forwarded \\
  & has HTML & n. of recipients \\
  & n. of ccd users & \\
  \textbf{Time characteristics:} & \multicolumn{2}{l}{hour of the day (24 features)} \\
  &  \multicolumn{2}{l}{day of the week (7 features)} \\
  \textbf{URL characteristics:} & \multicolumn{2}{l}{points to domain (variable \# of features)} \\
 \end{tabular}
 \caption{List of composition-habit features used in our approach.}
 \label{sec:featurescomposition}
\end{center}
\end{table*}

\noindent\textbf{Composition and sending habits.} Other habits that users develop regarding their email-sending behavior 
pertain to the way of composing emails. 
In the following, we describe this type of features. 
A complete list of composition-habit features can be found in
Table~\ref{sec:featurescomposition}.

\noindent\emph{1) Message characteristics (11 features).} These features capture specific
user habits in email composition. Examples of such habits are including
the original email at the end of a reply, including quotes to the original email
interleaved with the text, or adding a signature at the end of the email. 
Message-characteristic features are boolean, meaning that they are set to 1 if a
certain behavior is present in an email, and to 0 otherwise.

\noindent\emph{2) Time characteristics (31 features).} Users tend to send emails at specific
times of the day, and only during specific days. For example, most people
working in an office will send emails between 9 am and 5 pm, from Monday to
Friday. Given this observation, an email sent at midnight on a Saturday might be suspicious. 
These features keep information about when an email has been
sent. In particular, they look at the day of the week and at the hour at which
the email was composed. Similarly to other composition-habit features,
time-characteristic features are boolean. We define seven features for the days
of the week, and 24 features for the hours of the day.

\noindent\emph{3) URL characteristics (variable number of features).} Some users include URLs in their emails, 
such as links to web pages that are needed for their job, or to websites that
they consider interesting. Over time, the set of URL domains 
that a user includes in her emails tends to be limited~\cite{egele2013compa}. 
On the other hand, if the user sent an email with a URL pointing to a domain that she
has never included before, this might be considered as suspicious.

To instantiate URL-characteristic features, we need a set of domains $\textbf{L}_u$ that the
user, as well as other people in her organization, referenced in the past.
This helps identifying resources that are ``internal'' to the organization
(which should be referenced often in the company's emails).
We also include an ``\emph{other}'' category to take into account those domains
that were never referenced by anybody in the organization. Similarly to the other composition-habit features,
URL-characteristic features are boolean, and are set to 1 if that domain is
referenced in the email, and 0 otherwise.

\noindent\textbf{Interaction habits.}
The last type of features involves the social network of a user. Users
typically send a large deal of emails to a handful of contacts, usually coworkers or close
friends. Having an email sent to an address that was never contacted before
might thus contribute to the suspiciousness degree of an email, especially if the user under scrutiny does not usually
interact with many other users.

To characterize the social network of a user, we look at the
recipients email addresses (the \texttt{To:} field), as well
as at \emph{carbon copy} addresses (the \texttt{CC:}
field). We define four types of interaction-habit features, representing the
email addresses and domains that a user sends emails to. 
The \emph{recipient address list} features take into account the
email addresses that an email is addressed to, while the \emph{recipient domain
list} ones  look for the domains that those email addresses belong to. The idea
behind this distinction is that if a user sends an email to an address that she
has never referenced before, but that belongs to an organization that she often
interacts with, this is less suspicious than an email addressed to a completely
unknown domain. 
Similarly, we define a \emph{carbon copy address list} and a \emph{carbon copy
domain list} by analyzing the email addresses of the \texttt{CC:} field.

To instantiate the interaction-habit features, we need a list $\textbf{L}_a$ of email addresses
that the user, as well as the other people in the same organization, contacted in the
past. It is important to look at the email addresses that the user has never
contacted, but some of her coworkers have. This is because having a user sending
an email to an executive she has never contacted before is very suspicious, and
might be a sign of spearphishing. In addition, to account for those addresses
and domains with which nobody in the organization has interacted before, we
add, for each of the four feature types, an ``\emph{other}'' category.
Similarly, we leverage a list $\textbf{L}_d$ of domains to which the
users in the organization have written emails in the past. 

Interaction-habit features are boolean: they are set to 1 if an email is
addressed to the address (or domain) represented by a given feature, and to 0
otherwise. If, for any of the four feature types, all features of that type have a value of
0, the ``\emph{other}'' feature is set to 1.

\subsection{Building Users\\ Behavioral Profiles} \label{sec:building}

After extracting a feature vector for each email that a user sent, we leverage
them to build a behavioral model that is able to distinguish whether an email has likely been sent
by that user or not. 
To learn the distinguishing characteristics of the email-sending
behavior for a user $U$, \approach{} compares the feature vectors built from the emails
sent by the user ($\textbf{M}_u$) to the feature vectors built from a set of
legitimate emails sent by other people ($\textbf{M}_o$).
The challenge in picking $\textbf{M}_o$ is to select a set of emails that is
representative enough to make the most characteristic features of the behavior of
the user stand out.

Given a user $U$ who wrote a set of emails $\textbf{M}_u$, we pick the set of
emails $\textbf{M}_o$ as follows.
For each user $U_i$ in the organization (other than $U$), we keep
a set of emails that $U_i$ has sent in the past. We call this set
$\textbf{M}_{ui}$. In addition, we consider a ``special'' user $U_x$. The set of
emails $\textbf{M}_{ux}$ corresponding to the user $U_x$ consists of emails
that were
not written by the users in the organization. This set of emails could be a
subset of the emails that were received by the company's mail server, or a set
of publicly-available legitimate emails.
Then, for each email in $\textbf{M}_u$, we pick a random email written by
another user $U_i$ and add it to $\textbf{M}_o$. We change the user $U_i$ for
each email in $\textbf{M}_u$, in a round-robin fashion.
By doing this, we ensure that the distribution of emails written by different users in $\textbf{M}_o$ is uniform.

After having collected $\textbf{M}_u$ and $\textbf{M}_o$, 
\approach{}'s classifier is trained to learn the email-sending behavioral profile of user $U$.
To this end, we leverage Support Vector Machines (SVMs) trained
with Sequential Minimal Optimization (SMO)~\cite{platt1998sequential}. 
The SMO algorithm is an iterative algorithm used to efficiently solve the optimization problem required for training SVMs.
More details are provided in Section~\ref{sec:classifieranalysis}.

Since the email-sending behavior of a user is likely to slightly change over
time (for example, as the user makes new social connections), \approach{}
keeps updating the behavioral profile, by adding to the profile the new emails that the user
sends. The identity verification mechanism described in Section~\ref{sec:challenges}
ensures that the emails we add to the behavioral profile have
been genuinely written by the user.
Being able to constantly update a user's behavioral profile is important because it 
gets more accurate as the number
of emails sent by the user increases. 
However, the strength of the model also
depends on how consistent a user is in her email-sending habits. 
As we will discuss in Section~\ref{sec:classifieranalysis}, the features that we defined all
contribute in defining the email-sending behavior of a user. The weight of the different
features actually depends on each user's specific habits, and cannot be
generalized. 
In addition, some features are easier for an attacker to imitate
than others. For example, it is easy for an attacker to emulate the functional words that are most used by a user. However,
the more advanced style metrics, such as the \emph{Sichel measure}, are not as
easy to emulate. In any case, as we will show in Section~\ref{sec:adapting}, it
is difficult for an attacker to figure out which features he should imitate to evade
detection by our approach.

\section{Detecting Anomalous Emails} \label{sec:detecting}

After having built the email-sending behavioral profile for a user, our approach checks any
email that the user is sending against his profile. More specifically, our algorithm works as follows:

\noindent\textbf{Step 1:} For each email $M$ that user $U$ sends, we extract a feature vector $\textbf{V}_m$.

\noindent\textbf{Step 2:} We compare $\textbf{V}_m$ against the behavioral profile of $U$, which we
 call $\textbf{BP}_u$. If $\textbf{V}_m$ complies with $\textbf{BP}_u$, we
 validate the email as being written by $U$, and proceed to step 4. Otherwise, we
 consider $M$ as anomalous, and go to step 3.

\noindent\textbf{Step 3:} To verify that the email was written by the legitimate user $U$, we 
perform an identity verification. If $U$ correctly confirms her identity, $M$ is
 considered as a false positive, and we go to step 4. If $U$
 fails to confirm her identity (or decides not to, because she may recognize an ongoing 
 attack), the email is considered as malicious and is discarded. A notification may then be sent to an administrator for further investigation.
 In the next section, we describe how we envision the identity verification process to take place.

\noindent\textbf{Step 4:} We add $\textbf{V}_m$ to the set of feature vectors that are used to
 calculate $\textbf{BP}_u$. This information will be used the next time that the
 behavioral profile is updated.

It is not necessary to update the behavioral profile for a user for every
sent email. The reason is that, although the email-sending habits of a user change
over time, they do not change that fast. In addition, 
updating the behavioral profile for a user may require up to 30 seconds in the
current implementation.
For these reasons, we envision the
behavioral profile update as a batch process that could be performed daily or
weekly.

\subsection{Verifying a User's Identity.} \label{sec:challenges}
One of the main challenges that anti-spam systems have to
face are false positives. Flagging a legitimate email as spam has a high impact
on the user, because it might prevent her from seeing that email at all.
This is due to traditional anti-spam techniques operating on the receiving
side of the email process, where it is impossible to verify that the sender of an email is
who she actually claims to be. In contrast, operating on the sending side 
enables us to request the user to prove her identity when a 
certain email is looking suspicious, before emails are actually sent.

In our approach, we propose to perform an identity verification process by 
sending, for example, a confirmation code to a device controlled by the
user, and request the user to input that code as part of a two-factor
authentication process~\cite{aloul2009two}. 
This verification process might be a simple
method such as answering a security question or a more advanced
method, such as a text message sent to the user's mobile phone as part of a
two-factor authentication process~\cite{aloul2009two}.
Each method has advantages and disadvantages. 
However, analyzing the single
identity-verification methods that one could implement goes beyond the scope of this paper.
For our purposes, we just assume that by going through this process the user can
prove her identity with high confidence.

We are aware that having to go through an identity verification process might be an 
annoyance for users. However, there is always a trade-off that needs to be established between usability and security. 
So, we argue that if the number of validations that a user has to go through
is reasonably low, it is a fair price to pay to significantly increase the
security of a company. In Section~\ref{sec:classifieranalysis} we perform an
analysis by which we show that the number of identity verification processes required by
\approach{} is reasonably low, and probably acceptable for a user's perspective.

\section{Evaluation} \label{sec:evaluation}

In this section, we evaluate the effectiveness of \approach{}. 
First, we describe the evaluation datasets that we used in our experiments.
Then, we
perform an analysis of the classifier used to build the email-sending behavioral
profiles. We show that the behavioral profiles build by \approach{} are effective at
detecting attack emails sent by compromised accounts. 
Also, we analyze the resilience of our system to ``mimicry'' attacks 
and show how \approach{} is able to deal with and detect this type of
advanced attacks.

\subsection{Evaluation Datasets}

To evaluate \approach{} we leverage a number of email datasets. 
First, we leverage the \emph{Enron corpus}~\cite{enron} as a large dataset of legitimate emails. 
This publicly-available dataset contains the
emails sent by the executives of a large company over several years. 
The dataset comprises 148 users, accounting for 126,075 emails. 
The Enron dataset is representative of the type of
emails sent in a large corporation (in terms of sending times, language, interactions), and
this makes it suitable for our testing purposes. 
In the remainder of the paper, we call this dataset $\textbf{D}_1$.
As a second dataset of legitimate emails we use a set of emails that were
provided to a large security company by their customers for research purposes.
This dataset is made of 1,776 emails which we consider as useful
to complement $\textbf{D}_1$ because of their diversity. In particular,
they are useful to populate $\textbf{M}_{ux}$, as we explained in
Section~\ref{sec:building}. We call this dataset $\textbf{D}_2$.
We use the datasets $\textbf{D}_1$ and $\textbf{D}_2$ for training. In
particular, for each user in $\textbf{D}_1$, we build an email-sending behavioral profile, by
leveraging both the emails in $\textbf{D}_1$ and in $\textbf{D}_2$.

For testing purposes, we needed a number of emails sent from compromised accounts,
and preferably used as part of a targeted attack. The problem is that,
unlike regular spam, collecting a large amount of such emails is challenging.
To overcome this problem, we manually selected three datasets of malicious emails. These
emails come from a set of malicious messages detected by a large security company, 
which were submitted by their customers for manual analysis and validation.

The first dataset, that we call $\textbf{S}_1$, is composed of
generic spam emails. Such emails typically advertise goods or services, such as
stock trading, pharmaceuticals, and dating sites. The main difference between the
emails in $\textbf{S}_1$ and common spam is that a state-of-the-art system
failed in detecting them as malicious, and therefore we can consider them as
``hard'' to detect; we test \approach{} on this dataset to show that although
the system has not been designed to fight traditional spam, it performs well in
detecting it, in case it was sent by compromised email accounts. $\textbf{S}_1$ is composed of 43,274 emails.

The second dataset, that we call $\textbf{S}_2$, is composed of malicious emails
(mostly phishing scams) that were
sent by email accounts that had been compromised. We selected these
emails by looking at emails in $\textbf{S}_1$ that were malicious, but that had
valid \texttt{DKIM} and/or \texttt{SPF} records~\cite{Leiba2007,spf:06}. 
In total, $\textbf{S}_2$ contains 17,473 emails.

The third dataset, which we call $\textbf{S}_3$, is a dataset of more sophisticated spearphishing
emails. Such emails try to lure the user into handing out corporate-specific sensitive information
(such as access credentials, confidential documents, etc) to a malicious party, usually via social
engineering. 
As we said, spearphishing emails are particular insidious to
companies, because it can lead to high financial losses. $\textbf{S}_3$ contains 546 emails.
These emails went undetected by the defense systems deployed
by the security company, and were submitted by its customers after the attacks
had happened.
The emails in $\textbf{S}_2$ and $\textbf{S}_3$ closely resemble the threat model that we are trying to counter with \approach{}.
In the next sections, we leverage these datasets to evaluate the effectiveness of \approach{}. 
First, we investigate how representative of a user's behavior
the behavioral models built from $\textbf{D}_1$ and $\textbf{D}_2$ are.
Then, we leverage these behavioral models to see whether \approach{} would have
detected an anomaly, in case any of the users sent a malicious email from
$\textbf{S}_1$, $\textbf{S}_2$, or $\textbf{S}_3$.
As a last experiment, we investigate how easy would be to evade \approach{} by imitating a user's email-sending behavior. 
We do this by modifying the emails in $\textbf{S}_1$, $\textbf{S}_2$, and $\textbf{S}_3$ to
look more and more similar to each user's sending behavior.

\subsection{Analysis of the Classifier} \label{sec:classifieranalysis}

We start by describing how we selected the features used in
\approach{} to build behavioral user profiles. Then, we investigate how accurate these profiles are 
to determine the true authorship of an email.
Finally, we show that the writing habits are usually not sufficient to detect whether an email is
forged or not.

\noindent\textbf{Instantiation of the features.}
As we explained in Section~\ref{sec:features}, some of the features used by our
approach are specific to the organization in which the system is run. In
particular, we need to know which email addresses and domains have been
contacted previously by the users within an organization, as well as the domains
that have been referenced in the body of the emails, as part of the URLs. We leverage the dataset
$\textbf{D}_1$ to calculate the sets $\textbf{L}_u$, $\textbf{L}_a$, and $\textbf{L}_d$. For this particular dataset,
$\textbf{L}_u$ was composed of 595 domains, $\textbf{L}_a$ of 22,849
email addresses, and $\textbf{L}_d$ of 3,000 domains. Notice that, in a
production environment, the size of $\textbf{L}_u$, $\textbf{L}_a$, and
$\textbf{L}_d$ would increase over time, since the users in the company would
post more URLs, and contact new people. This means that the number of features
used by \approach{} increases over time as well. We argue that this is not a problem. Our experiments on $\textbf{S}_1$, which we omit for space reasons, show that the set of different domains that a user contacts in her emails over time grows slowly.

As for the writing-habit features, we needed to select a set of context-specific
words. We did this manually, by analyzing the most common words in the emails
of $\textbf{D}_1$ and picking those words that are
specific of the business of the company (i.e., finance, oil, and human
resources). In total, we selected 46 context-specific words, which are 
listed in Table~\ref{sec:featureswriting}.
We acknowledge that this process could be automated, 
but the manual selection worked well for our purposes, and previous
author-identification work used a similar approach~\cite{zheng2005framework}.

\begin{figure*}[t]
 \begin{subfigure}{0.49\textwidth}
  \includegraphics[scale=0.6]{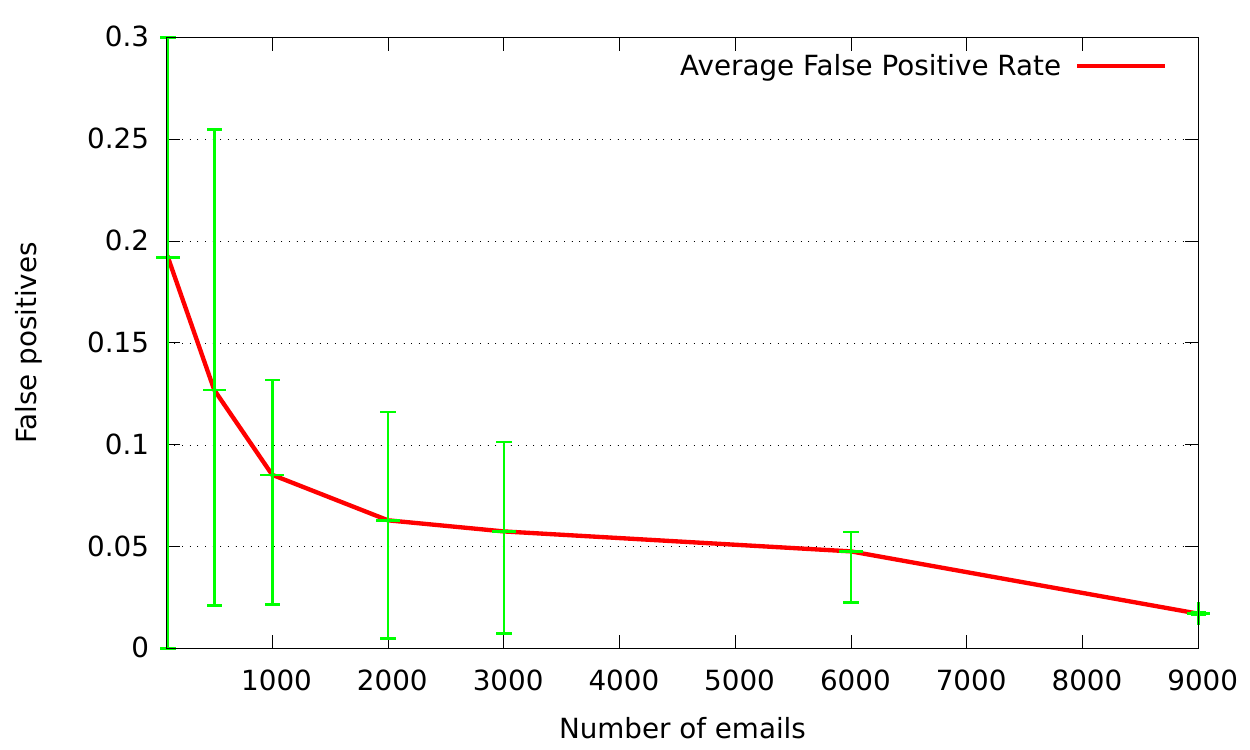}
  \caption{Analysis of the False Positive rate}
  \label{fig:falsepositives}
 \end{subfigure}
 \begin{subfigure}{0.49\textwidth}
  \includegraphics[scale=0.6]{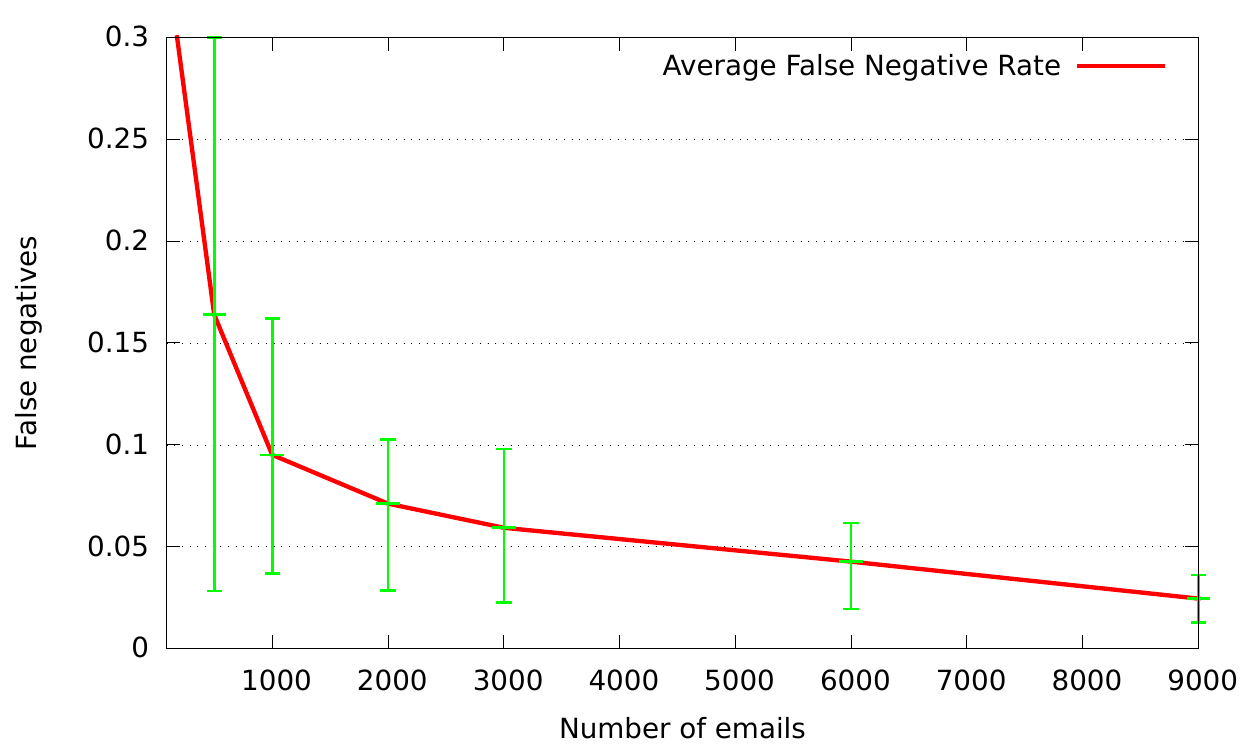}
  \caption{Analysis of the False Negative rate}
  \label{fig:falsenegatives}
 \end{subfigure}
 \caption{Analysis of false positives and false negatives on the ten-fold cross
 validation. The X-axis shows the number of emails that a user has sent in the
 past. As it can be seen, both false positives and false negatives decrease as
 the user's \emph{sent} email volume increases.}
\label{fig:accuracy}
\vspace*{-0.2in}
\end{figure*}

\noindent\textbf{Accuracy of the classifier.} 
To evaluate to what extent the \approach{} profiles are truly representative of the sending behavior of users, 
we proceeded as follows. First, for each user
$U$ in $\textbf{D}_1$, we extracted the sets $\textbf{M}_u$ and $\textbf{M}_o$ for that
user, following the algorithm described in Section~\ref{sec:building}. As said before, we use the emails sent by $U$ as positive
examples, and a mix of emails from $\textbf{D}_1$ and $\textbf{D}_2$ as negative examples.
In this experiment, we consider \approach{} to make a correct classification if
it attributes an email authored by a user $U$ to that user, and an incorrect
classification otherwise.

After having trained the system for each user, we performed a 10-fold cross
validation on them to investigate the accuracy of the behavioral profiles. 
The 10-fold cross validation gives us an idea of how the system would behave in
the wild, while encountering previously-unseen emails. In particular, it gives
us an estimate of how many emails would be incorrectly flagged as
malicious because of a change in behavior by a given user, as well as how many
attack emails would actually be missed by \approach{}.
In this experiment, a false
positive would indicate an email that was authored by the user, but flagged by
\approach{} as anomalous. In this case, an identity verification process would be started, 
by which the genuine user would have correctly confirmed her identity. 
We want the number of false positives to be low, because having
to confirm one's identity too often would become a users' annoyance. Conversely, a false
negative would indicate a forged email missed by \approach{}, thus mistakenly attributed to the legitimate user. 
We want false negatives to be as low as possible, since in a real scenario, each of them would
correspond to an attack that went undetected.

Intuitively, there are two factors that influence the robustness of a user's behavioral profile.
The first factor is the number of emails that a user has sent in the past.
Having a larger number of examples of a user's sending style and habits makes
the model more representative and less prone to false positives and false
negatives. The second factor is how consistent is the sending behavior of a user. 
A user always sending emails in the morning, to a limited set of 
recipients, will obviously be a lot more easily recognizable than a user who uses her account
for both professional and personal use and quite frequently sends emails at night.

The number of emails sent by users in $\textbf{D}_1$ varies substantially. On
average, every user in $\textbf{D}_1$ has sent 840 emails, with a standard deviation
of 1,345. The largest number of emails sent by a user in $\textbf{D}_1$ is 8,926.
The accuracy of \approach{} increases significantly as the number of emails sent
by a user increases, because the system can learn the typical behavior of that
user more accurately. 
It is challenging to show how the system behaves in a
figure, because any time the history of emails sent by the user increases, we
are evaluating a new system; for this reason, a Receiving Operating Curve (ROC)
is not suited to represent the accuracy of \approach{}. 
In Figure~\ref{fig:accuracy} we represent the average rate of false positives and
false negatives according to the \emph{sent} email volume of a user.
As Figure~\ref{fig:accuracy} shows, the accuracy of the email-sending behavioral
profile built by \approach{} increases as the user sends more emails. 
The error bars in the figure show that the accuracy of a behavioral profile does not only depend on
the email volume, but also on the user's style and habits. For users who have
consistent habits, \approach{} can achieve almost zero false positives and false
negatives. On the other hand, certain users having more variable habits end up
having higher rates of false positives and false negatives than the average. However, this
variability is reduced as the number of emails sent by the user increases.

Figure~\ref{fig:falsepositives} shows the average number of false positives generated during the
10-fold cross validation, broken down by the amount of emails sent in the past
by each user. As explained before, a false positive in this context would result in
the user being required to go through an identity verification mechanism. We note that, on average, 
a user who has sent at least 1,000
emails would have to confirm her identity for 1 in 12 emails. 
By increasing the sent email volume, a user who sent at least 8,000 emails would
have to confirm her identity on average for 1 in 58 emails that she sends. Given the average number of
emails that a typical corporate user sends nowadays --- 33 per day, 
according to a recent report~\cite{radicati}, reaching this amount of
interaction history would not take a long time. Moreover, these are average
numbers, thus users with a more stable email-sending behavior can already reach 2\% false
positives after having sent only 1,000 emails. These users would then have to go through the identity verification process for only 1 in every 50 emails that they send. 
We argue that these numbers are reasonably low and quite acceptable in a corporate environment, where the hassle of
confirming a user's identity is largely compensated by a significantly higher user protection against
identity and IP theft.

Similarly, Figure~\ref{fig:falsenegatives} shows the number of false negatives
for the 10-fold cross validation. As it can be seen, a sent history of 1,000 emails enables \approach{} to build a model able on average to block 90\% of the forged emails. 
Recall improves as the number of sent emails increases. The behavioral profile of a user who sent at least 8,000 emails has an average recall of 96\%.
A careful reader might notice that the accuracy of
\approach{} might be slightly under current state-of-the-art anti-spam systems.
However, as we previously said, the purpose of our system is very different from anti-spam techniques. 
We want to ensure that no malicious email is sent illegitimately on behalf of a user, and current anti-spam techniques were not designed to deal with such attacks.

\noindent\textbf{Analysis of the features.} Previous research showed that it is
possible to identify the author of an email just by looking at stylometric
features (what we refer to as \emph{writing habits} in this paper)~\cite{corney2003}.
However, Forsyth et al. showed that such approaches are only reliable in
the presence of a consistent amount of text~\cite{forsyth1996feature}. In
particular, they identified the minimal amount of text required for stylometry-based
author identification to become reliable (which is about 250 words). Unfortunately, 78\% of the emails in $\textbf{D}_1$ are under
this size limit. In particular, 50\% of the emails in that set are shorter than 100
words.

As we said, to deal with this issue of short email length, we use two other classes of features: 
\emph{composition habits} and \emph{interaction habits}.
We wanted to investigate the contribution of these features in detection accuracy, 
and confirm that writing-habit features alone are not sufficient.
To achieve this, we performed the 10-fold cross validation that we ran to
evaluate the classifier again, but this time we only used writing-habit features. The
results show that writing-habit features alone are indeed failing to obtain an
accurate detection. For a user with a 1,000 sent emails history, the average
number of false positives is now 22\% -- almost three times higher than for the full-fledged
classifier. The lowest rate of false positives obtained in this case is for users
having sent at least 8,000 emails, yet the FP-rate is still around 9.8\% -- almost six
times higher than the rate obtained with the full-fledged classifier. Clearly, while
stylometry-based methods might be useful in forensic cases, they are not sufficient
in this case to determine, with high confidence, whether an email has been sent by an attacker or not.

\subsection{Detecting Attack Emails} \label{sec:detecting}

\begin{figure}[h]
\includegraphics[width=0.46\textwidth]{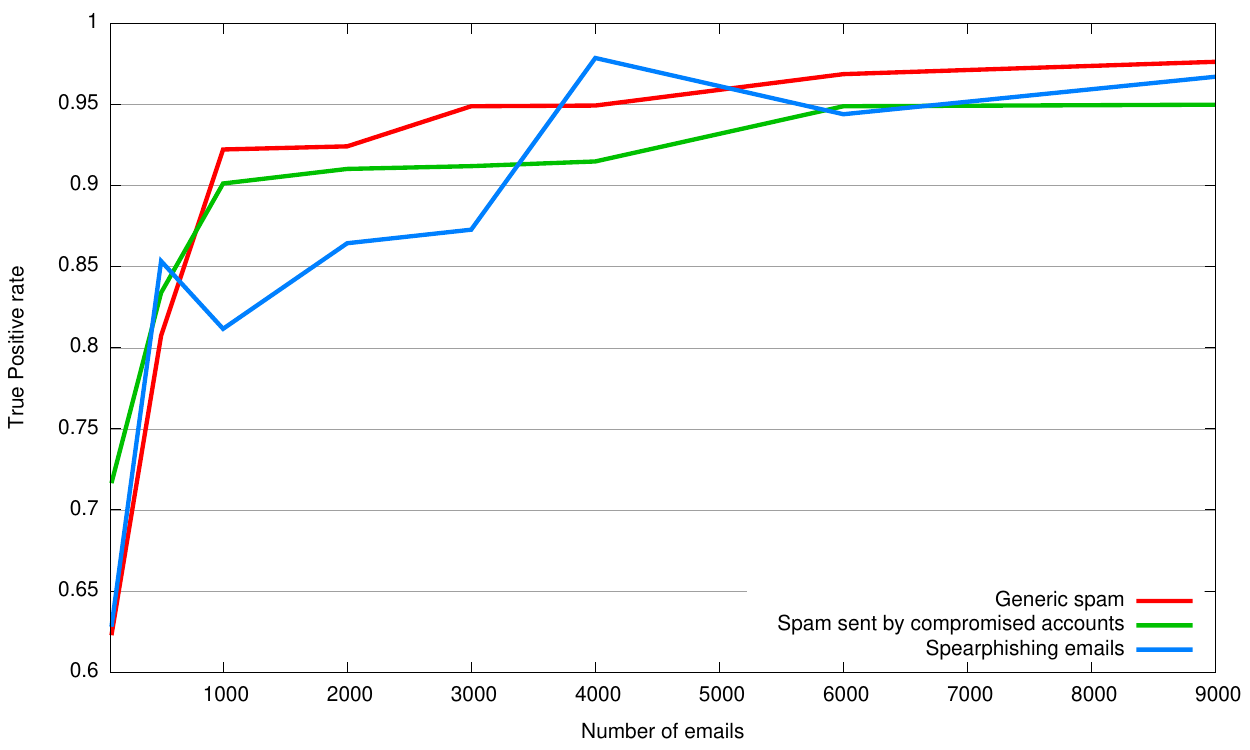}
\caption{Analysis of the True Positive rate for our attack datasets. On the X
axis we have the number of emails that a user sent in the past. As we can see,
\approach{} is able to block more and more emails as malicious as the sending
history of a user increases.}
\label{fig:attackplot}
\end{figure}

We now evaluate \approach{} on the attack datasets $\textbf{S}_1$, $\textbf{S}_2$,
and $\textbf{S}_3$. 
First, we created the email-sending behavioral profiles for each user $U$ in $\textbf{D}_1$, 
as explained in Section~\ref{sec:building}.
Then, for each email in $\textbf{S}_1$, $\textbf{S}_2$,
and $\textbf{S}_3$, and each user $U$, we edited the \texttt{From:} field in the email to look like it
was sent by $U$, and ran \approach{} against it, to see whether the email would
have been flagged as anomalous if it was being sent from $U$'s account. Since \approach{} does not use, 
at least at this stage, some header fields such as the \texttt{X-Mailer} or sender IP address, 
no additional editing was required for our test purposes.

Figure~\ref{fig:attackplot} shows the detection results of \approach{} on the
three datasets. 
As for the validation of the classifier, the
performance of \approach{} depends on how many emails each user has sent in the
past, as well as the consistency of a user's behavior while sending emails.
In general, an email history of 200 messages is enough to reach a true positive
rate of 80\%, while sent email logs of 1,000 emails or more lead to 90\% detection
rate. As a peak, \approach{} reaches 98\% true positives for certain users. 
These detection numbers are actually very promising and demonstrate very good performance of the system.
To put things into perspective, for the very same evaluation datasets $\textbf{S}_1$, $\textbf{S}_2$,
and $\textbf{S}_3$, most state-of-the-art systems would entirely fail to detect any of these emails as malicious. 
Hence, being able to detect most of these particularly insidious spearphishing emails 
is undoubtedly a major improvement over existing systems. In fact, \approach{} can be seen as an additional protection layer
that complements existing anti-spam systems, in order to block advanced spearphishing emails
that other email protection layers would not be able to detect.

\subsection{Fighting an Adapting Enemy} \label{sec:adapting}

\begin{table*}[t]
\begin{center}
\begin{tabular}{|l |c |c |c |c|}
\hline
\textbf{Type of Evasion} & \textbf{Failure} & \textbf{Success} & \textbf{No Effect} &
	\textbf{Avg. Change} \\
\hline
\texttt{C} & 5 & 0 & 143 & -0.3\% \\
     \texttt{T} & 81 & 63 & 4 & +3.4\% \\
\texttt{T + C} & 6 & 0 & 142 & -0.3\% \\
	\texttt{M\_10} & 87 & 49 & 12 & +2.4\% \\
 \texttt{M\_20} & 91 & 50 & 7 & +5.0\% \\
\texttt{T + TC} & 81 & 63 & 4 & +3.4\% \\
\texttt{T + TC + M\_10}& 73 & 73 & 2 & +5.7\% \\
\texttt{T + TC + M\_20} & 65 & 78 & 5 & +8.4\% \\
\hline
\end{tabular}
\caption{Summary of the results for different evasion strategies: Coworkers (C),
Time (T), Top contact (TC), Mimic 10 words (M\_10), Mimic 20 words (M\_20), and combinations of them.
This table shows that evading \approach{}'s detection is hard: even in the most
successful case of evasion, the attack fails for 43\% of the users.}
\label{tab:evasion}
\end{center}
\end{table*}

As we said previously, the techniques used in \approach{} for building the email-sending behavioral profiles 
enable us to extract
those characteristics that identify a certain user's sending behavior best, hence
performing accurate classification and obtaining better detection. Of course, an attacker could imitate a user's
sending behavior, as an attempt to evade detection by our system. 
An attacker who compromised a user's machine or email account typically has
access to the emails sent by that user in the past (for example, through
the \emph{Sent} folder of the user's mailer program). He can thus leverage these emails to
learn what are the most common characteristics in a users' email-sending
behavior, and replicate them in his malicious emails. However, since the
attacker does not have access to all emails sent by all other users in the
organization, it is difficult for him to know whether those characteristics, although
common in the user's behavior, distinguish the user well from the others.
In fact, the most common characteristics of a user's behavior might be shared with many 
other users, thus by replicating them an attacker would not obtain any
effect on the success of his attack. Even worse, he might focus on
characteristics that have a marginal importance or very low relevance in the behavioral profile built by
\approach{}, and make other characteristics stand out, in fact making detection
easier.

To investigate what would be the effects of an attacker actively trying to evade
our system, we developed a number of evasion techniques, and tested them on the
spearphishing dataset $\textbf{S}_3$. In particular, for each user $U$, and every
email $M$ in $\textbf{S}_3$, we extracted the feature vector $\textbf{V}_m$, and
modified a number of parameters, to make the email look more similar to the ones
typically sent by $U$. We then applied the same detection approach described in
Section~\ref{sec:detecting} to this modified dataset. In the
following, we describe the different evasion schemes that we developed.

\noindent\textbf{Coworkers.} This is the simplest evasion scheme. Instead of
sending attack emails to email addresses outside the user's organization,
the attacker sends an email to another user within 
the same company, with whom the victim exchanged at least one
email in the past. The destination address is picked at random.

\noindent\textbf{High activity time period.} In this evasion scheme, the attacker sends emails on the
day and hour during which the user has sent the highest number of emails in the
past.

\noindent\textbf{Top contact.} This technique is similar to the
\emph{coworkers} one, except that the destination address is chosen as the one to which
the user sent the highest number of emails.

\noindent\textbf{Mimic.} In this technique, the attacker tries to
replicate the writing style of the victim. In particular, he learns the $n$ most
common functional and context-specific words used by the user, and he uses them
in the attack emails with the same ratio as typically used by the victim in
her legitimate emails. We experimented two different evasion techniques of this
kind, replicating the 10 and 20 most common words used by a user, respectively.

We tested these evasion techniques described here above individually, as well as in conjunction with each other. 
Table~\ref{tab:evasion} provides a summary of the results for every (combination of) evasion techniques, compared with the
results obtained with the unmodified dataset. 
In each evasion scenario, the \emph{failure}, \emph{success}, and \emph{no
effect} columns indicate the number of victim users for which the evasion attack has failed, was successful, or did not change, respectively. 
The \emph{average change} field indicates the average percentage of emails that
successfully evaded \approach{} for that strategy, compared to the detection on
the unmodified dataset. A negative value indicates that, on average, \approach{}
performed better on the modified dataset than on the original one. 

As it can be seen, none of the evasion techniques guarantees that the attack emails will be
more successful in evading \approach{}. All evasion techniques (except for the
\emph{coworkers} and \emph{top contact} ones) provide, on average, an increase
in the number of emails that are not detected as malicious by our approach. However, even in the
most successful case, in which we use the \emph{time}, \emph{top contact}, and
\emph{mimic 20} techniques in conjunction, the evasion fails for 43\% of the
users, allowing \approach{} to detect more emails as malicious than for the
unmodified dataset. 
If we look at the entire user population, this advanced evasion scheme would have
enabled an additional 8.4\% of emails to evade detection; this means that,
for a user who sent 1,000 emails in the past, \approach{} would have still been
able to block 82\% spearphishing emails on average.
Given these results, we can conclude that it is not straightforward for 
an attacker to evade detection by \approach{}, and that our system is quite resilient against this type of evasion technique even when the
attacker tries to mimic the victim's email sending habits. 

An attacker might try a replay attack, by taking an email that was written
by a user in the past, and sending it again. This strategy has the limitation
that the attack is limited to the content that the user has already written in
the past. We acknowledge that in some cases this strategy might be
viable. However, this more extreme case of mimicry attack would be straightforward to 
mitigate, as we can easily trigger an identity verification process any
time an email results in a feature vector that is identical to another one observed in
the past.

Another technique that an attacker could try is imitating the email-sending
behavior learning phase of our system. To this end, he might leverage the emails that other users
sent to the victim in the past as $\textbf{M}_o$. 
The attacker can find these emails, for example, in the \emph{Inbox} of the victim's mailer program.
In principle, this technique
could help in making the attack more successful, and evade \approach{}. However,
the information that an attacker can learn from the emails received by a user in
the past is limited. For example, it does not give any information on
what the social networks of the other users in the company look like, and it
only shows the behavior that third party showed when interacting with that
specific user. An attacker might get additional knowledge of the company's
emails by compromising additional email accounts. If he obtained access
to a significant number of accounts, he might be able to replicate the learning process of
\approach{} and evade our system. However, an attack of such breadth is
hard to set up, and once an attacker gets such a pervasive presence is the company's
network, there is not much that our approach (or possibly any other) can do.

\section{Discussion and limitations} \label{sec:limitations}

Our results show that \approach{} is successful in detecting and blocking attack
emails that appear to have been written by a legitimate user, but have actually been
authored by an attacker abusing someone else's account. Like most detection systems, however, \approach{} has
some limitations.
The main limitation is that, to be effective, \approach{} requires an email
history of at least 1,000 emails. This makes it difficult to protect, for example, the
new hires of a company. We argue that email is such a pervasive communication
medium that it should not take long to obtain a sufficient number of emails even for
a new employee. In addition, a new hire is probably not going to be a good
target for an attacker, either due to a lack of visibility or because an attacker would prefer 
to target more influential people in the company.
These individuals, however, will have a long email-sending history, and \approach{} will
thus protect them effectively. Another possible limitation in a corporate setting is that
high-ranked executives might delegate their assistants to write some emails on
their behalf. This practice might generate false positives, because \approach{}
would detect that those emails were not written by the owner of the account. A
possible mitigation here is to learn multiple email-sending behaviors corresponding to a limited set 
of individuals who are using the same account, and thus avoid generating an alert if the email appears to
be authored by any of those users.

Another limitation of \approach{} is that writing-habit features 
are specific to the English language. If our approach had to protect the
employees of a company whose main language is different than English, we would
have to develop another set of language-specific features. Previous research
showed that this is feasible even for Asian languages, which have very
different characteristics than English~\cite{zheng2005framework}.

In Section~\ref{sec:adapting} we showed that it is difficult for attackers to
successfully evade our system. However, attackers could exploit weak points in
\approach{}'s deployment at specific companies. For example, if an organization
used a publicly-available set of emails as $\textbf{M}_{ux}$, an attacker might
get access to that dataset, use it in a similar way to learn the models 
and thus evade our system. However, the attacker would still not have access to the
emails used by all the other employees of the company, and the
knowledge of the attacker would still be incomplete. Similarly, an attacker
might try to build emails that resemble the victim's style, for example by using
a Context Free Grammar (CFG). I the model used by the attacker is
not complete, however, he will still not be guaranteed to succeed.

Another problem that we have to consider is the privacy of users. The email
sending behavior is built not only by leveraging a user's personal emails, but also by leveraging the
ones sent by her coworkers. However, feature vectors built from the email are kept among the client and the server,
and are never seen by the users. Also, the server has to only keep the feature
vector relative to an email, instead of the email itself. Therefore, we argue
that the privacy implications caused by \approach{} are still acceptable.

Another concern is that some domains, such as large webmail providers, have a very
diverse set of users, and thus it might be challenging to accurately model their behavior.
We argue that the focus of \approach{} is on corporate users, and we assume that their behavior is
more consistent than the one of general-purpose email providers. In addition,
large webmails have access to additional signals that are not included in our
threat model (such as login patterns and IP addresses), which can also be
leveraged to build a behavioral profile.

\section{Related Work}

Our approach protects the identity of users against attackers sending emails on
their behalf. To this end, we borrow some ideas from anti-spam techniques, as
well as from the field of forged text detection and authorship identification.
In the following, we
discuss how our approach is related to previous work, and elaborate on the
novelty of our method.

\noindent\textbf{Spam Filtering:} Existing work on spam filtering can be
distinguished in two main categories: \emph{origin-analysis} and \emph{content-analysis}
 techniques. Origin-analysis techniques try to determine whether
emails are good or bad by looking at their origin. Examples of characteristics that
are indicative of a malicious emails can be the IP address or autonomous system
that the email is sent from, or the geographical distance between the sender and
the
recipient~\cite{Hao:2009:SNARE,Ramachandran:07:behavioral,Stringhini:11:usenix,Venkataraman:2007:ENS}.
Other origin-based techniques include \emph{Sender Policy Framework} (SPF)~\cite{spf:06} and
\emph{DomainKeys Identified Mail} (DKIM)~\cite{Leiba2007}. These
techniques try to determine whether an email is actually coming from the address it
claims to come from, by looking at the sender IP, or at a signature in the email
headers. Origin-based techniques are widely deployed, because they allow servers 
to discard spam emails as soon as the malicious end connects to the mail server,
saving resources and time. In addition, they reach good coverage, because most
spam is sent by hosts that are part of a botnet, and therefore have a low
reputation~\cite{symantec}. However, in the scenario in which \approach{} works,
origin-based techniques are useless, because the only thing they can do is
confirming that an email has been sent by a certain account, regardless if it is a compromised one or not.

Content-analysis techniques look at the words in the message itself to determine
if it is spam or not. Proposed methods include Na\"ive Bayes, Support Vector
Machines, or other machine learning
algorithms~\cite{Drucker1999,Meyer2004,Sahami1998,Sculley:2007:ROS}. Other
systems detect spam by looking at malicious URLs in the
email~\cite{surbl,John:2009:SSB}. Content-analysis techniques work well in
detecting spam, however are too computationally intensive to be applied to every
email that a busy mail server receives~\cite{Taylor2006}. In \approach{}, we solve
this problem by analyzing emails as they get sent. We claim that this analysis
is feasible, because the amount of emails that a mail server sends is lower 
than the amount of emails that it receives. Another problem of
traditional content-analysis techniques is that they look for words that are
indicative of spam. In the presence of a targeted attack, there might be no such
words, since an attack email will use a language that is similar to the one used
in everyday business emails. This is why in \approach{} we learn the typical
sending behavior of a user and match it against the emails
she sends.

A number of systems have been proposed to counter specific types of spam, such
as phishing. Such systems either look at features in the attack emails that are
indicative of phishing content~\cite{fette2007learning}, or at characteristics of the
web page that the links in the email point to~\cite{zhang2007cantina}.
\approach{} is more general, since it can detect any type of attack emails that
is sent by compromised accounts. 
In addition, existing phishing techniques fail
in detecting those emails that rely on advanced social engineering tactics,
instead of redirecting the user to a phony login page.

Another category of spam detection techniques looks at the way in which spammers
use the TCP or SMTP protocol~\cite{kakavelakis2011,stringhini2012babel}. These
techniques work well in practice against most spam, but are focused on detecting hosts that belong
to a botnet, and are therefore useless in detecting the type of attacks that
\approach{} is designed to prevent.

\noindent\textbf{Email Forgery Detection:} A large corpus of research has been
performed on determining the authorship of written text. These techniques typically leverage
stylometry and machine learning and return the most probable author among a set of
candidates~\cite{calix2008stylometry,corney2003,iqbal2008novel,abbasi2008stylometric,frantzeskou2007identifying}.
From our point of view, these approaches suffer from two major problems: the first
one is that they typically need a set of possible authors, which in our case we
do not have. The second problem is that email bodies are often times too short 
to reliably determine the author by just looking at stylometry~\cite{forsyth1996feature}. Lin
et al. proposed a system that looks at the writing style of an email, and is
able to tell whether that email was written by an author or
not~\cite{linlightweight}. This approach may solve the first problem, but does not
solve the second one, in which we have emails that are too short to make a
meaningful decision. To mitigate this problem, in \approach{} we leverage
many other features other than stylometry, such as the times at which a user
sends emails, or her social network.

Khonji et al. presented ASCAI~\cite{khonji2011mitigation}, a system that detects
if an email author has likely been forged. ASCAI looks at the
most common n-grams in a user's emails, and flag as anomalous emails that
contain words that the user rarely uses. Unlike \approach{}, ASCAI looks for any
word, instead of focusing on writeprint features (such as functional words). For
this reason, this system would fail in detecting spearphishing emails whose
content is about the same topics that the user typically discusses, but that
have been authored by a different person. 
\approach{}, on the other hand, has
been designed to detect this type of stealthy spearphishing emails, and is
therefore effective in blocking them.

Stolfo et al. presented the Email Mining Toolkit
(EMT)~\cite{stolfo2003behavior,stolfo2006behavior}. This tool mines email logs
to find communities of users who frequently interact with each other. After learning the
communities, the system flags as anomalous emails that are addressed to people
outside them. Although EMT leverages an idea similar to \approach{}'s
interaction features, it is tailored at detecting large-scale threats, such as
worms spreading through email. The fact that \approach{} leverages other types
of features allow our system to detect more subtle, one-of-a-kind attack emails.  

Egele et al. proposed a system that learns the behavior of users on Online 
Social Networks (OSN) and flags anomalous messages as possible
compromises~\cite{egele2013compa}. Because of the high number of false
positives, their system can only detect large-scale campaigns, by
aggregating similar anomalous messages. As we have shown, \approach{} is able to
detect attacks that are composed of a single email, and which have not been seen
before.

\section{Conclusions}

We presented \approach{}, a system that protects the mailbox of corporate users by checking whether an email has been written
by the legitimate owner of an email account. This work is the first step towards the
protection of individuals and companies against advanced email attacks, such as
spearphishing. \approach{} is able to learn the typical sending behavior of
the account's owner and can subsequently check all emails sent from the account against this
profile in order to block advanced spearphishing attacks sent from a compromised email account. 
By performing experiments on real world datasets, we also showed that \approach{} can effectively block attacks that state-of-the-art
protection systems are unable to detect, and that an attacker has no clear strategy to make
his emails look legitimate in order to evade our detection system.

\section*{Acknowledgments}

This work was supported by the Symantec Research Labs Graduate Fellowship for 
the year 2012. The authors would like to thank the people at Symantec, in 
particular Marc Dacier, David T. Lin, Dermot Harnett, Joe Krug, David Cawley, and Nick Johnston for their support and comments. We would also like to thank Adam Doup\`e and Ali Zand for reviewing an early version of this paper. Your comments 
were highly appreciated.

\bibliographystyle{abbrv} 
\small{
\bibliography{biblio} 

\begin{thebibliography}{10}

\bibitem{spamhaus}
{The Spamhaus Project}.
\newblock \url{http://www.spamhaus.org}.

\bibitem{abbasi2008stylometric}
A.~Abbasi, H.~Chen, and J.~F. Nunamaker.
\newblock Stylometric identification in electronic markets: Scalability and
  robustness.
\newblock {\em Journal of Management Information Systems}, 2008.

\bibitem{afroz2012detecting}
S.~Afroz, M.~Brennan, and R.~Greenstadt.
\newblock Detecting hoaxes, frauds, and deception in writing style online.
\newblock In {\em IEEE Symposium on Security and Privacy}, 2012.

\bibitem{aloul2009two}
F.~Aloul, S.~Zahidi, and W.~El-Hajj.
\newblock Two factor authentication using mobile phones.
\newblock In {\em IEEE/ACS International Conference on Computer Systems and
  Applications}, 2009.

\bibitem{calix2008stylometry}
K.~Calix, M.~Connors, D.~Levy, H.~Manzar, G.~MCabe, and S.~Westcott.
\newblock Stylometry for e-mail author identification and authentication.
\newblock {\em Proceedings of CSIS Research Day, Pace University}, 2008.

\bibitem{corney2003}
M.~W. Corney.
\newblock {Analysing E-mail Text Authorship for Forensic Purposes}.
\newblock 2003.

\bibitem{Drucker1999}
H.~Drucker, D.~Wu, and V.~N. Vapnik.
\newblock {Support vector machines for spam categorization}.
\newblock In {\em IEEE transactions on neural networks}, 1999.

\bibitem{egele2013compa}
M.~Egele, G.~Stringhini, C.~Kruegel, and G.~Vigna.
\newblock {COMPA: Detecting Compromised Social Network Accounts}.
\newblock In {\em Symposium on Network and Distributed System Security (NDSS)},
  {2013}.

\bibitem{fette2007learning}
{Fette, I. and Sadeh, N. and Tomasic, A.}
\newblock {Learning to Detect Phishing Emails}.
\newblock 2007.

\bibitem{forsyth1996feature}
R.~Forsyth and D.~Holmes.
\newblock {Feature finding for text classification}.
\newblock In {\em {Literary and Linguistic Computing}}, {1996}.

\bibitem{frantzeskou2007identifying}
G.~Frantzeskou, E.~Stamatatos, S.~Gritzalis, C.~E. Chaski, and B.~S. Howald.
\newblock Identifying authorship by byte-level n-grams: The source code author
  profile (scap) method.
\newblock {\em International Journal of Digital Evidence}, 2007.

\bibitem{Hao:2009:SNARE}
S.~Hao, N.~A. Syed, N.~Feamster, A.~G. Gray, and S.~Krasser.
\newblock {Detecting Spammers with SNARE: Spatio-temporal Network-level
  Automatic Reputation Engine}.
\newblock In {\em USENIX Security Symposium}, 2009.

\bibitem{iqbal2008novel}
F.~Iqbal, R.~Hadjidj, B.~Fung, and M.~Debbabi.
\newblock A novel approach of mining write-prints for authorship attribution in
  e-mail forensics.
\newblock {\em {Digital Investigation}}, 2008.

\bibitem{jagatic2007social}
T.~N. Jagatic, N.~A. Johnson, M.~Jakobsson, and F.~Menczer.
\newblock Social phishing.
\newblock {\em Communications of the ACM}, 2007.

\bibitem{John:2009:SSB}
J.~P. John, A.~Moshchuk, S.~D. Gribble, and A.~Krishnamurthy.
\newblock {Studying Spamming Botnets Using Botlab}.
\newblock In {\em USENIX Symposium on Networked Systems Design and
  Implementation (NSDI)}, 2009.

\bibitem{kakavelakis2011}
G.~Kakavelakis, R.~Beverly, and Y.~J.
\newblock {Auto-learning of SMTP TCP Transport-Layer Features for Spam and
  Abusive Message Detection}.
\newblock In {\em USENIX Large Installation System Administration Conference},
  2011.

\bibitem{khonji2011mitigation}
M.~Khonji, Y.~Iraqi, and A.~Jones.
\newblock Mitigation of spear phishing attacks: A content-based authorship
  identification framework.
\newblock In {\em Internet Technology and Secured Transactions (ICITST), 2011
  International Conference for}, 2011.

\bibitem{enron}
B.~Klimt and Y.~Yang.
\newblock {Introducing the Enron Corpus}.
\newblock In {\em CEAS}, 2004.

\bibitem{Leiba2007}
B.~Leiba.
\newblock {DomainKeys Identified Mail (DKIM): Using digital signatures for
  domain verification}.
\newblock In {\em CEAS}, 2007.

\bibitem{linlightweight}
E.~Lin, J.~Aycock, and M.~Mannan.
\newblock Lightweight client-side methods for detecting email forgery.
\newblock In {\em {Workshop on Information Security Applications (WISA)}},
  {2012}.

\bibitem{nightdragon}
{McAfee Inc.}
\newblock {Global Energe Cyberattacks: ``NightDragon''}, 2011.

\bibitem{Meyer2004}
T.~Meyer and B.~Whateley.
\newblock {SpamBayes: Effective open-source, Bayesian based, email
  classification system}.
\newblock In {\em CEAS}, 2004.

\bibitem{narayanan2012feasibility}
A.~Narayanan, H.~Paskov, N.~Z. Gong, J.~Bethencourt, E.~Stefanov, E.~C.~R.
  Shin, and D.~Song.
\newblock On the feasibility of internet-scale author identification.
\newblock In {\em IEEE Symposium on Security and Privacy}, 2012.

\bibitem{parmar2012protecting}
B.~Parmar.
\newblock Protecting against spear-phishing.
\newblock {\em Computer Fraud \& Security}, 2012.

\bibitem{platt1998sequential}
J.~Platt et~al.
\newblock Sequential minimal optimization: A fast algorithm for training
  support vector machines.
\newblock 1998.

\bibitem{Ramachandran:07:behavioral}
A.~Ramachandran, N.~Feamster, and S.~Vempala.
\newblock {Filtering Spam with Behavioral Blacklisting}.
\newblock In {\em ACM Conference on Computer and Communications Security
  (CCS)}, 2007.

\bibitem{Sinha:2010:ISB}
M.~B. S.~Sinha and F.~Jahanian.
\newblock {Improving Spam Blacklisting Through Dynamic Thresholding and
  Speculative Aggregation}.
\newblock In {\em Symposium on Network and Distributed System Security (NDSS)},
  2010.

\bibitem{Sahami1998}
M.~Sahami, S.~Dumais, D.~Heckermann, and E.~Horvitz.
\newblock {A Bayesian approach to filtering junk e-mail}.
\newblock {\em Learning for Text Categorization}, 1998.

\bibitem{Sculley:2007:ROS}
D.~Sculley and G.~M. Wachman.
\newblock {Relaxed Online SVMs for Spam Filtering}.
\newblock In {\em ACM SIGIR Conference on Research and Development in
  Information Retrieval}, 2007.

\bibitem{tibet}
{Securelist}.
\newblock {Android Trojan Found in Targeted Attack}.
\newblock
  \url{http://www.securelist.com/en/blog/208194186/Android_Trojan_Found_in_Targeted_Attack},
  2013.

\bibitem{stolfo2006behavior}
S.~J. Stolfo, S.~Hershkop, C.-W. Hu, W.-J. Li, O.~Nimeskern, and K.~Wang.
\newblock Behavior-based modeling and its application to email analysis.
\newblock {\em ACM Transactions on Internet Technology (TOIT)}, 2006.

\bibitem{stolfo2003behavior}
S.~J. Stolfo, S.~Hershkop, K.~Wang, O.~Nimeskern, and C.-W. Hu.
\newblock Behavior profiling of email.
\newblock In {\em Intelligence and Security Informatics}. 2003.

\bibitem{stringhini2012babel}
G.~Stringhini, M.~Egele, A.~Zarras, T.~Holz, C.~Kruegel, and G.~Vigna.
\newblock {B@BEL: Leveraging Email Delivery for Spam Mitigation}.
\newblock In {\em USENIX Security Symposium}, 2012.

\bibitem{Stringhini:11:usenix}
G.~Stringhini, T.~Holz, B.~Stone-Gross, C.~Kruegel, and G.~Vigna.
\newblock {BotMagnifier: Locating Spambots on the Internet}.
\newblock In {\em USENIX Security Symposium}, 2011.

\bibitem{symantec}
{Symantec Corp.}
\newblock Symantec intelligence report.
\newblock
  \url{http://www.symanteccloud.com/mlireport/SYMCINT_2013_01_January.pdf},
  2013.

\bibitem{Taylor2006}
B.~Taylor.
\newblock {Sender reputation in a large webmail service}.
\newblock In {\em CEAS}, 2006.

\bibitem{radicati}
{The Radicati Group}.
\newblock {Email Statistics Report}.
\newblock
  \url{http://www.radicati.com/wp/wp-content/uploads/2011/05/Email-Statistics-Report-2011-2015-Executive-Summary.pdf}.

\bibitem{thonnard2012industrial}
O.~Thonnard, L.~Bilge, G.~O'Gorman, S.~Kiernan, and M.~Lee.
\newblock Industrial espionage and targeted attacks: understanding the
  characteristics of an escalating threat.
\newblock In {\em Symposium on Recent Advances in Intrusion Detection (RAID)},
  2012.

\bibitem{worm}
{Threatpost}.
\newblock {New Email Worm Turns Back the Clock on Virus Attacks}.
\newblock
  \url{http://threatpost.com/en_us/blogs/new-email-worm-turns-back-clock-virus-attacks-090910},
  2010.

\bibitem{trend2012spearphishing}
{Trend Micro Inc.}
\newblock {Spear-Phishing Email: Most Favored APT Attack Bait}, 2012.

\bibitem{tweedie1998variable}
F.~Tweedie and R.~Baayern.
\newblock {How variable may a constant be? Measures of lexical richness in
  perspective}.
\newblock {\em Computers abd the humanities}, 1998.

\bibitem{Venkataraman:2007:ENS}
S.~Venkataraman, S.~Sen, O.~Spatscheck, P.~Haffner, and D.~Song.
\newblock {Exploiting Network Structure for Proactive Spam Mitigation}.
\newblock In {\em USENIX Security Symposium}, 2007.

\bibitem{spf:06}
M.~Wong and W.~Schlitt.
\newblock {RFC 4408: Sender Policy Framework (SPF) for Authorizing Use of
  Domains in E-Mail, Version 1}.
\newblock \url{http://tools.ietf.org/html/rfc4408}, 2006.

\bibitem{yule1944statistical}
G.~Yule.
\newblock {The statistical study of literary vocabulary}.
\newblock {\em Cambridge University Press}, 1944.

\bibitem{surbl}
M.~Zalewski.
\newblock {p0f v3}.
\newblock \url{http://lcamtuf.coredump.cx/p0f3/}, 2012.

\bibitem{zhang2007cantina}
{Zhang, Y. and Hong, J.I. and Cranor, L.F.}
\newblock {Cantina: a Content-based Approach to Detecting Phishing Web Sites}.
\newblock 2007.

\bibitem{zheng2005framework}
R.~Zheng, J.~Li, H.~Chen, and Z.~Huang.
\newblock {A Framework for Authorship Identification of Online Messages:
  Writing-Style Features and Classification Techniques}.
\newblock {\em Journal of the American Society for Information Science and
  Technology}, 2005.

\end{thebibliography}
}

\end{document}